\documentclass[prl,aps,showpacs]{revtex4}
\usepackage{bm}
\begin{document}

%
%
%
\title{Diffusive mass transfer by non equilibrium fluctuations: Fick's law revisited}

\author{Doriano Brogioli$^1$, and Alberto Vailati$^2$}

\affiliation{$^1$Dipartimento di Fisica and Istituto Nazionale per
la Fisica della Materia (INFM), Universit\`a degli Studi di
Cagliari Strada provinciale Monserrato-Sestu km 0,7, 09042
Monserrato (CA), Italy}

\affiliation{$^2$Dipartimento di Fisica and Istituto Nazionale per
la Fisica della Materia (INFM), Universit\`a degli Studi di
Milano, via Celoria 16, 20133 Milano, Italy}

\date{\today}

\begin{abstract}
Recent experimental and theoretical works have shown that giant
fluctuations are present during diffusion in liquid systems. We
use linearized fluctuating hydrodynamics to calculate the net mass
transfer due to these non equilibrium fluctuations. Surprisingly
the mass flow turns out to coincide with the usual Fick's one. The
renormalization of the hydrodynamic equations allows us to
quantify the gravitational modifications of the diffusion
coefficient induced by the gravitational stabilization of long
wavelength fluctuations.
\end{abstract}


\maketitle

 Recent experiments have
shown that giant non equilibrium concentration fluctuations are a
universal feature of diffusion in binary liquid mixtures and
macromolecular solutions \cite {vailati97_2,brogioli00}. The
presence of long-range correlations in a diffusing liquid mixture
is a surprising result, as diffusion is usually thought to give
rise to a homogeneous mixing at the molecular level, other length
scales playing no role in the process. The results of these
experiments have been interpreted by means of an extension of
Landau's Fluctuating Hydrodynamics  \cite {Landau} to
time-dependent diffusion processes \cite {vailati98}.

Linearized hydrodynamics shows that the fluctuations are generated
by thermal velocity fluctuations, due to the presence of a
macroscopic non equilibrium concentration gradient. The mechanism
is analogous to that first predicted \cite {law89_2,segre93_3} and
then observed \cite {segre93,li94,vailati96,li98} for stationary
thermal diffusion in a binary liquid mixture. Velocity
fluctuations parallel to the macroscopic gradient displace parcels
of fluid into regions with different concentration, thus giving
rise to concentration fluctuations \cite{weitz97}. Since the
fluctuations displace mass along the concentration gradient, one
might wonder if they contribute significantly to the macroscopic
diffusive mass flow.

We will show that, quite unexpectedly, the net mass transfer due
to non equilibrium fluctuations corresponds to the macroscopic
Fick's mass flow. Traditionally the diffusive flow has been
interpreted as due to the random hops of the molecules in the
fluid. Here we show that Fick's flow is due to random hops of
parcels of fluid whose dimensions span all the length scales from
the molecular up to the macroscopic one. This is a surprising
result, as usually thermal fluctuations are thought of as small
perturbations of a macroscopic state, while here we show that the
macroscopic flow is determined by a second-order effect in the
fluctuations. The introduction of new degrees of freedom, the
hydrodynamic fluctuations, determines an ultraviolet divergence of
the diffusion coefficient, which we regularize by imposing a
cut-off at large wave vectors. The regularization procedure
involves the introduction of an arbitrary length scale. By
imposing that the cut-off length scale is of the order of the
correlation length of the mixture we find that the diffusion
coefficient displays the usual Stokes-Einstein form and is of the
same order of magnitude of the phenomenological one. However the
introduction of an arbitrary length scale into the theory can be
avoided by a renormalization of the hydrodynamic equations. The
renormalization procedure allows us to quantify the modifications
of the phenomenological diffusion coefficient due to the
gravitational stabilization of long wavelength fluctuations.

We will now derive the net contribution of the non equilibrium
fluctuations to the mass transfer. The sample considered is a
binary mixture of concentration $c\left(\mathbf{x},t\right)$ under
rest.
 A local fluctuation in the current of mass can be written
as the second order product

\begin{equation}
\label {flutt_j} \delta \mathbf{j}_f(\mathbf{x},t)=\rho \delta
c(\mathbf{x},t) \delta \mathbf{u}(\mathbf{x},t),
\end{equation}

where  $\delta c(\mathbf{x},t)$ is the local concentration
fluctuation and $\delta \mathbf{u}(\mathbf{x},t)$ the local
velocity fluctuation.

The net contribution of the fluctuations to the diffusive current
is obtained by averaging over the configurations of the system:

\begin{equation}
\label {flusso_spaziale_j}\mathbf{j}_f(\mathbf{x},t)=\rho
\left<\delta c(\mathbf{x},t) \delta
\mathbf{u}(\mathbf{x},t)\right>.
\end{equation}

Equation (\ref {flutt_j}) shows that a current fluctuation takes
place whenever a concentration fluctuation and a velocity
fluctuation occur simultaneously at the same place. At
thermodynamic equilibrium fluctuations in the velocity and
concentration are not correlated. Therefore the probability of a
current fluctuation is small and with random sign, and the net
flow $\mathbf{j}_f$ averages to zero. Under non equilibrium
conditions, however, velocity fluctuations are correlated to
concentration fluctuations \cite {vailati98,law89_2,segre93_3}.
The presence of a macroscopic concentration gradient determines
the existence of velocity-induced concentration fluctuations: a
velocity fluctuation determines a small vortex inside the fluid
and this brings parcels of fluid with a certain concentration into
layers of fluid with a different concentration. On the contrary,
the gravity force determines the opposite mechanism: whenever a
concentration fluctuation is produced, the buoyancy force tends to
displace the fluctuation. Therefore terms in the average of Eq.
(\ref {flusso_spaziale_j}) are correlated and this determines a
non vanishing mass flow $\mathbf{j}_f$ .

We will shortly derive an explicit expression for $\mathbf{j}_f$
by determining the cross-correlation properties of concentration
and velocity fluctuations in the presence of a macroscopic
concentration gradient. This task can be accomplished by using
Landau's linearized hydrodynamics \cite{Landau}. Basically one
writes down the relevant hydrodynamic equations for a binary
mixture. The equations are then linearized for small fluctuations
of the hydrodynamic variables around their macroscopic values and
random source terms are added to them.

The equations thus obtained for a binary mixture where a uniform
concentration gradient is present are \cite {vailati98}

\begin{equation}
\label {teo_eq_linear_fourier} \left\{
\begin{array}{l}
\delta c\left(\mathbf{q},\omega\right) \left(i\omega+D_0
q^2\right) = -\delta \mathbf{u}\left(\mathbf{q},\omega\right)
\cdot \bm{\nabla}c +i \mathbf{q} \cdot \mathbf{F}
\left(\mathbf{q},\omega\right)\\
 \delta
\mathbf{u}\left(\mathbf{q},\omega\right) \left(i\omega+\nu
q^2\right) = \beta \delta c\left(\mathbf{q},\omega\right)
\left(\mathbf{g}-\mathbf{q}\frac{\mathbf{q}\cdot\mathbf{g}}{q^2}\right)
+\frac{i}{\rho} \mathbf{q} \wedge \mathbf{S}
\left(\mathbf{q},\omega\right)
\end{array}
\right.,
\end{equation}

where $D_0$ is the bare diffusion coefficient, $\nu$ the kinematic
viscosity, $\mathbf{g}$ the gravitational acceleration,
$\beta=\rho^{-1}\left(\partial \rho / \partial c \right)$ and
$c\left(\mathbf{q},\omega\right)$  is the space-time Fourier
transform for the concentration fluctuations defined by

\begin{equation}
\label {teo_def_fourier} \delta c\left(\mathbf{q},\omega\right) =
\frac{1}{\left(2 \pi\right)^4} \int{e^{\displaystyle
i\mathbf{q}\cdot\mathbf{x} +i\omega t} \delta
c\left(\mathbf{x},t\right) \mathrm{d}\mathbf{x} \mathrm{d}t},
\end{equation}

and an analogous expression holds for the velocity fluctuations.

In Eq.
(\ref{teo_eq_linear_fourier})$\mathbf{F}(\mathbf{q},\omega)$ and
$\mathbf{S}(\mathbf{q},\omega)$ represent random source currents
for concentration fluctuations and velocity fluctuations,
respectively. As customary, their correlation properties are
assumed to retain their equilibrium values \cite {cohen71}:

\begin{equation}
\label {teo_correlazione_FF} \left<F_i(\mathbf{q},\omega)
F_j(\mathbf{q}',\omega')\right> = \delta_{ij}
\delta(\mathbf{q}+\mathbf{q}') \delta(\omega+\omega') \frac{K_B
T}{8\pi^4}\frac{D_0}{\rho}\left(\frac{\partial c}{\partial
\mu}\right)_{p,T},
\end{equation}

\begin{equation}
\label {teo_correlazione_SS} \left<S_{i}(\mathbf{q},\omega)
S_{j}(\mathbf{q}',\omega')\right> =
\delta_{ij}\delta(\mathbf{q}+\mathbf{q}')
\delta(\omega+\omega')\frac{K_B T}{8\pi^4}\nu\rho,
\end{equation}

\begin{equation}
\label {teo_correlazione_FS} \left<F_{i}(\mathbf{q},\omega)
S_{j}(\mathbf{q}',\omega')\right> = 0.
\end{equation}

Let us briefly comment Eq. (\ref {teo_eq_linear_fourier}). The
left terms specify that concentration  and velocity fluctuations
decay with  lifetimes $1/D_0q^2$ and $1/\nu q^2$, respectively.
The right terms are source terms for the fluctuations. Beyond the
random currents $\mathbf{F}(\mathbf{q},\omega)$ and
$\mathbf{S}(\mathbf{q},\omega)$, they contain the cross-coupling
between velocity and concentration fluctuations. The
cross-coupling term in the concentration equations shows that a
velocity fluctuation $\delta \mathbf{u}$ couples with the
macroscopic concentration gradient $\bm{\nabla} c$, thus giving
rise to a concentration fluctuation. The cross coupling term in
the velocity equation represents the acceleration acting on a
concentration fluctuation due to the buoyancy force. We have
assumed that the fluid is incompressible; therefore the
acceleration acting on the fluctuation is the component of gravity
perpendicular to the wave vector $\mathbf{q}$.

The explicitation of  $\delta c$ and $\delta\mathbf{u}$ in Eq.
(\ref {teo_eq_linear_fourier}) yields

\begin{equation}
\label {flux_sol_c} \delta c\left(\mathbf{q},\omega\right)=\frac{
i\left(i\omega+\nu
q^2\right)\mathbf{q}\cdot\mathbf{F}\left(\mathbf{q},\omega\right)
-\frac{i}{\rho}\mathbf{q}\wedge\mathbf{S}\left(\mathbf{q},\omega\right)
\cdot\bm{\nabla}c }{ \left(i\omega+D_0 q^2\right)\left(i\omega+\nu
q^2\right)+ \beta
\left(\mathbf{g}-\mathbf{q}\frac{\mathbf{q}\cdot\mathbf{g}}{
q^2}\right) \cdot \bm{\nabla} c }
\end{equation}

and

\begin{eqnarray}
\delta \mathbf{u}\left(\mathbf{q},\omega\right)=\frac{
i\mathbf{q}\cdot\mathbf{F}\left(\mathbf{q},\omega\right)
\beta\left(\mathbf{g}-\mathbf{q}\frac{\mathbf{q}\cdot\mathbf{g}}{
q^2}\right) +\frac{i}{\rho}\left(i\omega+D_0
q^2\right)\mathbf{q}\wedge
\mathbf{S}\left(\mathbf{q},\omega\right)}{ \left(i\omega+D_0
q^2\right)\left(i\omega+\nu q^2\right)+ \beta
\left(\mathbf{g}-\mathbf{q}\frac{\mathbf{q}\cdot\mathbf{g}}{
q^2}\right) \cdot \bm{\nabla} c}
\nonumber
\\
\label {flux_sol_u} + \frac{\frac{i}{\rho}\frac{\beta}{i\omega+\nu
q^2} \left[
\mathbf{q}\wedge\mathbf{S}\left(\mathbf{q},\omega\right)
\bm{\nabla} c \cdot
\left(\mathbf{g}-\mathbf{q}\frac{\mathbf{q}\cdot\mathbf{g}}{
q^2}\right) -
\mathbf{q}\wedge\mathbf{S}\left(\mathbf{q},\omega\right) \cdot
\bm{\nabla} c
\left(\mathbf{g}-\mathbf{q}\frac{\mathbf{q}\cdot\mathbf{g}}{
q^2}\right) \right] }{ \left(i\omega+D_0
q^2\right)\left(i\omega+\nu q^2\right)+ \beta
\left(\mathbf{g}-\mathbf{q}\frac{\mathbf{q}\cdot\mathbf{g}}{
q^2}\right) \cdot \bm{\nabla} c}.
\end{eqnarray}

We are now in the position to evaluate the  mass flow $\mathbf
{j}_f$. In the reciprocal space Eq. (\ref {flusso_spaziale_j})
becomes

\begin{equation}
\label {flusso_trasf} \mathbf{j}_f= \rho \int{\left<\delta
c(\mathbf{q},\omega)\delta \mathbf{u}(\mathbf{q}',\omega') \right>
\mathrm{d} \mathbf{q} \mathrm{d} \mathbf{q}'\mathrm{d} \omega
\mathrm{d} \omega'}.
\end{equation}

The integrand  in Eq. (\ref {flusso_trasf}) can be decomposed as
the sum of terms containing correlations of the random currents
$\mathbf {F}$ and $\mathbf {S}$. Due to (\ref
{teo_correlazione_FS}) the only non vanishing terms are those
originated from self-correlations:

\begin{equation}
\label {flusso_decomp} \left<\delta c(\mathbf{q},\omega)\delta
\mathbf{u}(\mathbf{q}',\omega')\right>= \left<\delta
c(\mathbf{q},\omega)\delta
\mathbf{u}(\mathbf{q}',\omega')\right>_{F}+ \left<\delta
c(\mathbf{q},\omega)\delta
\mathbf{u}(\mathbf{q}',\omega')\right>_{S},
\end{equation}

where

\begin{equation}
\label {flux_eq_corr_cu_F} \left<\delta c(\mathbf{q},\omega)\delta
\mathbf{u}(\mathbf{q}',\omega')\right>_{F}=
\delta\left(\mathbf{q}+\mathbf{q}'\right)\delta\left(\omega+\omega'\right)
\frac{K_B T D_0}{8 \pi^4 \rho} \left(\frac{\partial c}{\partial
\mu}\right)_{p,T} \beta\frac{ q^2\left(i\omega+\nu q^2\right)
\left(\mathbf{g}-\mathbf{q}\frac{\mathbf{g}\cdot\mathbf{q}}{
q^2}\right) }{\left|\left(i\omega+D_0 q^2\right)\left(i\omega+\nu
q^2\right)+ \beta
\left(\mathbf{g}-\mathbf{q}\frac{\mathbf{q}\cdot\mathbf{g}}{
q^2}\right) \cdot \bm{\nabla}c\right|^2}
\end{equation}

and

\begin{equation}
\label {flux_eq_corr_cu_S} \left<\delta c(\mathbf{q},\omega)\delta
\mathbf{u}(\mathbf{q}',\omega')\right>_{S}=
-\delta\left(\mathbf{q}+\mathbf{q}'\right)\delta\left(\omega+\omega'\right)
\frac{K_B T \nu}{8 \pi^4 \rho} \frac{ q^2\left(i\omega+D_0
q^2\right)
\left(\bm{\nabla}c-\mathbf{q}\frac{\bm{\nabla}c\cdot\mathbf{q}}{
q^2}\right) }{\left|\left(i\omega+D_0 q^2\right)\left(i\omega+\nu
q^2\right)+ \beta
\left(\mathbf{g}-\mathbf{q}\frac{\mathbf{q}\cdot\mathbf{g}}{
q^2}\right) \cdot \bm{\nabla}c\right|^2}.
\end{equation}

The denominator of Eqs. (\ref {flux_eq_corr_cu_F}-\ref
{flux_eq_corr_cu_S}) is an even function of $\omega$; therefore in
the evaluation of the integral in Eq. (\ref{flusso_trasf}) we can
neglect odd functions of $\omega$ in the numerator.

By combining  Eqs. (\ref {flusso_trasf}-\ref {flux_eq_corr_cu_S})
and by assuming that $\mathbf{\hat{z}} = \mathbf{g}/g =
\bm{\nabla}c/\left|\nabla c\right|$, the current $\mathbf{j}_f$
due to non equilibrium fluctuations is finally determined:

\begin{equation}
\label {flux_flusso_calc} \mathbf{j}_f=-\rho
D_f\left[\bm{\nabla}c-\beta\mathbf{g} \left(\frac{\partial
c}{\partial \mu}\right)_{p,T}\right],
\end{equation}

where

\begin{equation}
\label {D_primo} D_f=\frac{K_B T \nu D_0}{8 \pi^4 \rho}
\int{\frac{q^4
\left(\mathbf{\hat{z}}-\mathbf{q}\frac{\mathbf{\hat{z}}\cdot\mathbf{q}}{
q^2}\right)\cdot\mathbf{\hat{z}} }{\left|\left(i\omega+D_0
q^2\right)\left(i\omega+\nu q^2\right)+ \beta g \nabla c
\left(\mathbf{\hat{z}}-\mathbf{q}\frac{\mathbf{\hat{z}}\cdot\mathbf{q}}{
q^2}\right) \cdot
\mathbf{\hat{z}}\right|^2}\mathrm{d}\mathbf{q}\mathrm{d}\omega}.
\end{equation}

Surprisingly the flow $\mathbf{j}_f$ displays the customary
dependence of Fick's flow on the driving force $\bm{\nabla}c$ and
on the barodiffusion flow.

Equation (\ref {D_primo}) can be rewritten by factorizing its
denominator as
$\left(\omega^2+\omega_+^2\right)\left(\omega^2+\omega_-^2\right)$
and by approximating the two roots as
$\omega^2_+=D_0^2q^4\left\{1+\left[q/q_{RO}\left(g\right)\right]^4
\left(\mathbf{\hat{z}}-\mathbf{q}\frac{\mathbf{\hat{z}}\cdot\mathbf{q}}{
q^2}\right)\cdot\mathbf{\hat{z}}\right\}^2$ and
$\omega^2_-=\nu^2q^4$. Basically this approximation holds true for
$\nu >>D$, a condition fulfilled by most binary liquid mixtures;
further details are discussed in Ref. \cite {vailati98}:

\begin{equation}
\label {D_primo_1} D_f=\frac{K_B T}{\left(2 \pi\right)^3 \rho
\nu}\int{\frac{ q^2
\left(\mathbf{\hat{z}}-\mathbf{q}\frac{\mathbf{\hat{z}}\cdot\mathbf{q}}{
q^2}\right)\cdot\mathbf{\hat{z}} }{q^4+q_{RO}\left(g\right)^4
\left(\mathbf{\hat{z}}-\mathbf{q}\frac{\mathbf{\hat{z}}\cdot\mathbf{q}}{
q^2}\right)\cdot\mathbf{\hat{z}} }\mathrm{d}\mathbf{q}},
\end{equation}

where

\begin{equation}
\label {teo_def_qro} q_{RO}\left(g\right)=\sqrt[4]{\frac{g \beta
\nabla c}{\nu D_0}}
\end{equation}

is a roll-off wave vector due to the gravity force \cite
{segre93_3}. As thoroughly discussed in Ref. \cite {vailati98},
the roll-off wave vector marks the transition from the diffusive
relaxation of fluctuations at large wave vectors to the
gravitational stabilization of fluctuations at smaller ones. At
the roll-off wave vector the diffusional and gravitational
fluctuation relaxation times coincide.

By introducing polar coordinates Eq. (\ref {D_primo_1}) becomes

\begin{equation}
\label {D_primo_2} D_f^Q=\frac{K_B T}{\left(2 \pi\right)^3 \rho
\nu} \int_0^Q{\mathrm{d}q \int_0^{\pi}{\mathrm{d}\theta
\int_0^{2\pi}{\mathrm{d}\varphi
\frac{q^4\sin^3\left(\theta\right)}{q^4+q_{RO}\left(g\right)^4\sin^2\left(\theta\right)}
}}}.
\end{equation}

The integral on $q$ is divergent. Therefore we have regularized it
by introducing a brute-force cut-off at the arbitrary wave vector
$Q$.
 The evaluation of the integrals in $\varphi$ and $q$ yields

\begin{equation}
\label {D_primo_3}  D_f^Q=\frac{K_B T}{\left(2 \pi\right)^2 \rho
\nu} \int_0^{\pi}{\mathrm{d}\theta
Q\left[1-\frac{\sqrt{2}\pi}{4}\frac{q_{RO}}{Q}\sqrt{\sin\theta}+
O\left(\frac{q_{RO}\sqrt{\sin\theta}}{Q}\right)^4
\right]\sin^3\theta}.
\end{equation}

By assuming that $q_{RO}\left(g\right)/Q \ll 1$ Eq. (\ref
{D_primo_3}) becomes

\begin{equation}
\label {D_primo_5} D_f^Q\left(g\right)=\frac{4}{3}\frac{K_B
T}{\left(2 \pi\right)^2\rho\nu}Q -\sqrt{2}I\frac{\pi}{4}\frac{K_B
T}{\left(2 \pi\right)^2\rho\nu}q_{RO}\left(g\right) + Q
O\left(q_{RO}\left(g\right)/Q\right)^4,
\end{equation}

where

\begin{equation}
I=\int_0^{\pi}{\mathrm{d}\theta \sin^{7/2}\theta} \approx
1.248599~.
\end{equation}

In the absence of gravity $q_{RO}=0$, and Eq.(\ref {D_primo_5})
becomes
\begin{equation}
\label {D_primo_4} D_f^Q=\frac{2}{3}\frac{K_B T}{\pi\rho\nu
\Lambda},
\end{equation}

where $\Lambda=2\pi/Q$ can be roughly assumed to be of the order
of correlation length of the fluid. It should be noticed that the
diffusion coefficient $D_f$ of Eq. (\ref {D_primo_4}) does not
depend on the bare mass diffusion coefficient $D_0$ which appeared
in the linearized fluctuating hydrodynamic equations (\ref
{teo_eq_linear_fourier}) and (\ref {teo_correlazione_FF}).

Quite remarkably, Eq. (\ref {D_primo_4}) exhibits the usual
Stokes-Einstein dependence on temperature, viscosity and
correlation length, the proportionality factor  being of the right
order of magnitude. This result, combined with Eq.(\ref
{flux_flusso_calc}), strongly suggests that the mass transfer due
to fluctuations  and Fick's mass flow can be identified. This is a
striking result, as it shows for the first time that the diffusive
mass transfer is entirely determined by non equilibrium
fluctuations occurring at length scales
 ranging from the microscopic up to the macroscopic one.

Although it is physically reasonable to assume that wave vectors
larger than a typical molecular one have no meaning, the
imposition of a cut-off involves the introduction of an arbitrary
length scale, which in turn requires to introduce a microscopic
model for the molecular interactions. This necessity can be
eliminated by a renormalization of the linearized hydrodynamic
equations. The normalization procedure we will discuss in the
following closely mirrors the renormalization of mass of Quantum
Field Theory \cite{lebellac} and of Classical Electrodynamics
\cite {feynman}.
 The overall diffusive mass flow is the superposition of
the contribution due to fluctuations $\mathbf{j}_f$ and the bare
fickean mass flow:

\begin{equation}
\label {j_overall} \mathbf{j}\left(g\right) =
\mathbf{j}_f^Q\left(g\right) + \mathbf{j}_0^Q\left(g\right) =
 - \rho \left(
D_f^Q\left(g\right)+ D_0^Q\right) \left[\bm{\nabla}c - \beta
\mathbf{g} \left(\frac{\partial c}{\partial
\mu}\right)_{p,T}\right].
\end{equation}

The renormalization condition is

\begin{equation}
\label {renorm1} \mathbf{j}\left(g=0\right) = - \rho D
\bm{\nabla}c,
\end{equation}

where $D$ is the phenomenological diffusion coefficient in the
absence of gravity.

By combining Eq. (\ref {j_overall}) and Eq. (\ref {renorm1}) the
renormalization condition can be rewritten as

\begin{equation}
\label {renorm3eq} D_0^Q = D - D_f^Q\left(g=0\right),
\end{equation}

and, by inserting Eqs. (\ref {D_primo_5},\ref {renorm3eq})  into
Eq. (\ref {j_overall}),
 the renormalized mass current is

\begin{equation}
\label {def_flusso_fenomenologico} \mathbf{j}_R\left(g\right) = -
\rho D_R\left(g\right)\left[\bm{\nabla}c - \beta \mathbf {g}
\left(\frac{\partial c}{\partial \mu}\right)_{p,T}\right],
\end{equation}

where $D_R$ is the renormalized diffusion coefficient

\begin{equation}
\label {D_rinorm_g} D_R\left(g\right) = D -
\sqrt{2}\frac{\pi}{4}I\frac{K_B T}{\left(2
\pi\right)^2\rho\nu}q_{RO}\left(g\right).
\end{equation}

In Eq. (\ref {D_rinorm_g}) the $Q$ dependence has been eliminated
by letting $Q \to \infty$, because $D_R$ does not diverge. The
microscopic details have been reabsorbed into $D$, the only
experimental parameter needed by the theory in order to obtain the
value of the diffusion coefficient in the presence of gravity.

Equation  (\ref {D_rinorm_g}) shows that the diffusion coefficient
is smaller due to the presence of gravity, the additional term
having the Stokes-Einstein form,  the only  length scale involved
being $2\pi/q_{RO}$.

This modification of the diffusion coefficient is due to the fact
that, in the presence of gravity, fluctuations with wave vector
smaller than $q_{RO}$ are stabilized and therefore do not
contribute to the mass transfer. Therefore $q_{RO}$ acts as an
infrared cut-off for the fluctuations. As the length scales
stabilized by gravity are large, the relative correction
introduced by gravity is very small, being of the order of
$q_{RO}\xi$, where we have assumed a Stokes-Einstein dependence of
the phenomenological diffusion coefficient $D$ on the correlation
length  $\xi$ of the fluid. For an ordinary liquid mixture
$q_{RO}$ can be as high as $10^3~\mathrm{cm}^{-1}$ and $\xi$ is of
the order of $10^{-7}~\mathrm{cm}$. Therefore the gravitational
correction is estimated to be roughly one part in $10^4$.
 In the case of critical fluids and macromolecular solutions the
 correlation length can grow very large, thus suggesting that the
 gravitational correction to the diffusion coefficient can get large.
  However, in order for our hydrodynamic
description to be valid, the correlation length has to be
significantly smaller than $1/q_{RO}$. Therefore the largest
correction to the diffusion coefficient reliably predicted by this
description can be roughly estimated to be of the order of 10\%.
Such a variation in the diffusion coefficient should be observable
by performing microgravity experiments on critical fluids.

The authors wish to thank Marzio Giglio for stimulating
discussion. Work partially supported by the Italian Space Agency
(ASI).

\end{document}